\pdfoutput=1
\documentclass[a4paper,11pt]{article}
\usepackage{amsmath,amssymb,amsfonts,mathtools}
\allowdisplaybreaks[4]
\usepackage[noconfig]{refstyle}
\usepackage[dvipsnames]{xcolor}
\usepackage[hyperfootnotes=false, linktocpage=true, colorlinks, citecolor=blue, linkcolor=blue, urlcolor=Maroon]{hyperref}
\usepackage{subfigure}
\usepackage[bf]{caption}
\usepackage[numbers,sort&compress]{natbib}
\usepackage{array,tabularx,booktabs,longtable}
\usepackage{geometry}
\usepackage[normalem]{ulem}

\let\eqref=\relax
\newref{eq}{name={eq.~},Name={Eq.~},names={eqs.~},Names={Eqs.~},rngtxt={-},refcmd=(\ref{#1})}
\newref{f}{name={Footnote~},Name={Footnote~},names={Footnotes~},Names={Footnotes~}}
\newref{app}{name={Appendix~},Name={Appendix~},names={Appendixes~},Names={Appendixes~}}
\newref{tab}{name={Table~},Name={Table~},names={Tables~},Names={Tables~}}
\newref{ch}{name={Chapter~},Name={Chapter~},names={Chapters~},Names={Chapters~}}
\newref{sec}{name={Section~},Name={Section~},names={Sections~},Names={Sections~}}
\newref{fig}{name={Figure~},Name={Figure~},names={Figures~},Names={Figures~}}
\numberwithin{equation}{section}
\newcommand{\nn}{\nonumber}
\newcommand{\be}{\begin{equation}}
\newcommand{\ee}{\end{equation}}
\newcommand{\bea}{\begin{equation}\begin{aligned}}	% note: abbreviations for \begin{align} and \end{align} don't work!
\newcommand{\eea}{\end{aligned}\end{equation}}		% note: \begin{equation}\begin{split}... produces pdf/hyperref warnings:
\newcommand{\iddots}{\mathinner{\mkern2mu\raise1pt\hbox{.}\mkern2mu \raise4pt\hbox{.}\mkern2mu\raise7pt\hbox{.}\mkern1mu}}

\providecommand{\id}{\leavevmode\hbox{\small$\mathrm{1}$\kern-3.8pt\normalsize$\mathrm{1}$}}
\def\fnote#1#2{\begingroup\def\thefootnote{#1}\footnote{#2}
     \addtocounter{footnote}{-1}\endgroup}
\newcommand{\varstr}[2]{\vrule height #1 depth #2 width0pt}

\begin{document}
\title{{$~$\\ \Large \bf Hypercharge Flux in Heterotic Compactifications\\[12pt]}}
\author{
Lara B. Anderson${}^{1}$,
Andrei Constantin${}^{2,3}$,
Seung-Joo Lee${}^{1,4}$,
Andre Lukas${}^{2}$
}

\date{}
\maketitle
\begin{center} {\vskip -1mm\small  
       ${}^1${\it Department of Physics, Robeson Hall, Virginia Tech \\ Blacksburg, VA 24061, U.S.A.}\\[0.2cm]
       ${}^2${\it Rudolf Peierls Centre for Theoretical Physics, Oxford University,\\
       $~~~~~$ 1 Keble Road, Oxford, OX1 3NP, U.K.\\[0.2cm]
       ${}^3${\it Department of Physics and Astronomy, Uppsala University, \\ Box 520, SE-751 20 Uppsala, Sweden}\\[0.2cm]
       ${}^4${\it School of Physics, Korea Institute for Advanced Study, Seoul 130-722, Korea}}}\\[40pt]
       
\fnote{}{lara.anderson@vt.edu}
\fnote{}{andrei.constantin@physics.uu.se}
\fnote{}{seungsm@vt.edu}
\fnote{}{lukas@physics.ox.ac.uk}

\end{center}

\begin{abstract}
\noindent We study heterotic Calabi-Yau models with hypercharge flux breaking, where the visible $E_8$ gauge group is directly broken to the standard model group by a non-flat gauge bundle, rather than by a two-step process involving an intermediate grand unified theory and a Wilson line. It is shown that the required alternative $E_8$ embeddings of hypercharge, normalized as required for gauge unification, can be found and we classify these possibilities. However, for all but one of these embeddings we prove a general no-go theorem which asserts that no suitable geometry and vector bundle leading to a standard model spectrum can be found. Intuitively, this happens due to the large number of index conditions which have to be imposed in order to obtain a correct physical spectrum in the absence of an underlying grand unified theory.
\end{abstract}

\thispagestyle{empty}
\setcounter{page}{0}
\newpage

%\tableofcontents

\section{Introduction}
Particle physics model building in the context of the $E_8\times E_8$ heterotic string \cite{Gross:1984dd, Gross:1985fr, Gross:1985rr, Candelas:1985en} on smooth Calabi-Yau manifolds~(see \cite{Braun:2009qy, Braun:2011ni, Blumenhagen:2005ga, Braun:2005ux, Braun:2005bw, Braun:2005nv, Bouchard:2005ag, Blumenhagen:2006ux, Blumenhagen:2006wj, Anderson:2007nc, Anderson:2008uw, Anderson:2009mh, Buchbinder:2014qda, Buchbinder:2014sya, Buchbinder:2013dna, He:2013ofa} for example) has, traditionally, been based on intermediate grand unified theories (GUTs), typically with gauge group $SU(5)$ or $SO(10)$, and subsequent GUT breaking by a Wilson line. A crucial benefit of this approach is the relative ease with which a quasi-realistic particle spectrum can be obtained~--~a~single index condition has to be imposed at the GUT level in order to guarantee three chiral families and this chiral asymmetry is preserved by the Wilson line.

In this paper, we study a different model-building approach where the visible $E_8$ gauge group is broken to the standard model group directly by flux, without an intermediate GUT and a Wilson line. This approach, based on direct flux breaking, is popular in the context of F-theory models (see Refs.~\cite{Beasley:2008kw, Donagi:2008kj,Buican:2006sn} and also the recent progress in \cite{Dudas:2010zb,Mayrhofer:2013ara,Braun:2014pva, Heckman:2013sfa}) but has also been considered in the heterotic context~\cite{Blumenhagen:2006ux,Blumenhagen:2006wj,Blumenhagen:2005ga} (including in the context of heterotic orbifolds, e.g.~\cite{Dienes:1995sq}).

Here, we study such models systematically. We assume that the non-Abelian part of the standard model group is embedded into $E_8$ via the maximal subgroup $SU_W(2)\times SU_c(3)\times SU(6)\subset E_8$. Further, we assume that the structure group of the (visible) bundle resides in $SU(6)$ and is of the general split type $S(U(n_1)\times\dots\times U(n_f))\subset SU(6)$, where $n_1+\dots +n_f=6$. In this case, the low-energy gauge group is $SU_W(2)\times SU_c(3)\times S(U(1)^f)$ and hypercharge has to be embedded into $S(U(1)^f)$.  

After setting out the general structure of these models in the next section, we study their detailed properties in two steps. In section~\ref{yemb} we focus on group-theoretical aspects. Specifically, we classify the possible embeddings of hypercharge which can lead to a viable physical spectrum and which have the correct normalization to be consistent with the standard picture of gauge unification.
In section~\ref{nogo}, we analyse the underlying Calabi-Yau geometries and bundles which might lead to such models. 

\section{Basic structure of models}
To begin, we describe the general structure of the models we consider in this paper. We study compactifications of the $E_8\times E_8$ heterotic string on Calabi-Yau manifolds with holomorphic, poly-stable bundles. We require that the two $E_8$ factors remain hidden from each other, so the entire standard model group and all standard model multiplets should originate from one $E_8$ only. The non-Abelian part of the standard model group is embedded into this $E_8$ factor via the sub-group chain $SU_W(2)\times SU_c(3)\subset SU_W(2)\times SU_c(3)\times SU(6)\subset E_8$ while hypercharge, $U_Y(1)$, resides in the $SU(6)$ factor. The conventional model-building approach would be to embed the whole standard model group into $E_8$ via an intermediate (GUT) $SU(5)$ group. Here, we do not demand this - after having fixed the embedding of $SU_W(2)\times SU_c(3)$ as described above, the embedding of hypercharge is left arbitrary.

The structure group, $H$, of the bundle in the observable sector can be at most $SU(6)$ since we require that $SU_W(2)\times SU_c(3)$ is contained in the commutant of $H$ within $E_8$. Here, we study the different possible unitary splittings of this maximal structure group; that is, we consider structure groups
\begin{equation}
 H=S(U(n_1)\times\cdots \times U(n_f))\;,\quad \sum_{a=1}^fn_a=6\; ,\eqlabel{Hdef}
\end{equation} 
which are classified by the partitions of $6$. There are $11$ such partitions which, in terms of the vector ${\bf n}=(n_1,\ldots ,n_f)$, are given by
\begin{equation}
\begin{aligned}
 {\bf n}=&(6),\;(5,1),\;(4,2),\;(3,3),\;(4,1,1),\;(3,2,1),\;(2,2,2),\;\\
 &(3,1,1,1),\;(2,2,1,1),\;(2,1,1,1,1),\;(1,1,1,1,1,1)\; . \eqlabel{splits}
\end{aligned}
\end{equation}  
The low-energy gauge group is then
\begin{equation}
 G=SU_W(2)\times SU_c(3)\times S(U(1)^f)~, \eqlabel{Gdef}
\end{equation} 
with hypercharge embedded into $S(U(1)^f)$, in a way that will be specified later. The additional $U(1)$ symmetries in \eqref{Gdef} will generically be Green-Schwarz anomalous with associated super-massive gauge bosons. However, for specific choices they can be non-anomalous and massless, a situation which we need to engineer for hypercharge. We will state the required conditions for this in the next section. To work out the possible low-energy multiplets, we start with the branching
\begin{equation}
 {\bf 248}_{E_8}\rightarrow \left[({\bf 3},{\bf 1},{\bf 1})\oplus ({\bf 1},{\bf 8},{\bf 1})\oplus ({\bf 1},{\bf 1},{\bf 35})\oplus ({\bf 1},{\bf 3},\overline{\bf 15})
 \oplus ({\bf 1},\overline{\bf 3},{\bf 15})
 \oplus ({\bf 2},{\bf 3},{\bf 6})\oplus ({\bf 2},\overline{\bf 3},\overline{\bf 6})\oplus ({\bf 2},{\bf 1},{\bf 20})\right]\eqlabel{248dec}
\end{equation} 
of the adjoint of $E_8$ into $SU_W(2)\times SU_c(3)\times SU(6)$ representations. To further decompose the $SU(6)$ representations into representations of $H$, we introduce the notation ${\cal F}_a$ and $Adj_a$ for the fundamental and adjoint of $SU(n_a)$, respectively, and also write $S(U(1)^f)$ charges as vectors ${\bf q}=(q_1,\ldots ,q_f)$ subject to the identification ${\bf q}\sim {\bf q}'\Leftrightarrow {\bf q}-{\bf q}'\in \mathbb{Z}{\bf n}$. With this notation, the $SU(6)$ representations in the decomposition~\eqref{248dec} further decompose as follows
\begin{equation}\eqlabel{SU6dec}
 \begin{array}{lll}
  {\bf 6}&\rightarrow&\bigoplus_{a=1}^f({\cal F}_a)_{{\bf e}_a}\\[5pt] 
  \overline{\bf 6}&\rightarrow&\bigoplus_{a=1}^f(\bar{\cal F}_a)_{-{\bf e}_a}\\[5pt]
  {\bf 15}&\rightarrow&\bigoplus_{a=1}^f(\wedge^2{\cal F}_a)_{2{\bf e}_a}\oplus\bigoplus_{a<b}({\cal F}_a\otimes{\cal F}_b)_{{\bf e}_a+{\bf e}_b}\\[5pt]
  \overline{\bf 15}&\rightarrow&\bigoplus_{a=1}^f(\wedge^2\bar{\cal F}_a)_{-2{\bf e}_a}\oplus\bigoplus_{a<b}(\bar{\cal F}_a\otimes\bar{\cal F}_b)_{-{\bf e}_a-{\bf e}_b}\\[5pt]
{\bf 20}&\rightarrow&\bigoplus_{a=1}^f(\wedge^3{\cal F}_a)_{3{\bf e}_a}\oplus\bigoplus_{a\neq b}(\wedge^2{\cal F}_a\otimes {\cal F}_b)_{2{\bf e}_a+{\bf e}_b}\oplus\bigoplus_{a<b<c}({\cal F}_a\otimes{\cal F}_b\otimes{\cal F}_c)_{{\bf e}_a+{\bf e}_b+{\bf e}_c} \\[5pt]
 {\bf 35}&\rightarrow&\bigoplus_{a=1}^f(Adj_a)_{{\bf 0}}\oplus\bigoplus_{a\neq b}({\cal F}_a\otimes \bar{\cal F}_b)_{{\bf e}_a-{\bf e}_b}\; ,
\end{array}
\end{equation}
where the subscript denotes the $S(U(1)^f)$ charge and $\bold e_a$ are the six-dimensional standard unit vectors. To parametrize the embedding of hypercharge into $S(U(1)^f)$, we introduce a vector
\begin{equation}
 {\bf y}=(y_1,\ldots ,y_f)\eqlabel{ydef}\; ,\qquad {\bf n}\cdot{\bf y}=0\; ,
\end{equation} 
such that $Y(F)={\bf y}\cdot{\bf q}(F)$ is the hypercharge of a multiplet $F$ with $S(U(1)^f)$ charge ${\bf q}(F)$. 

\vspace{8pt}
The vector bundle with the required structure group $H$ has the general form
\begin{equation}
 V=\bigoplus_{a=1}^f U_a\; ,\eqlabel{Vdef}
\end{equation}
where $U_a$ is a rank $n_a$ bundle with structure group $U(n_a)$ and we require that $c_1(V)=\sum_a c_1(U_a)\stackrel{!}{=}0$. The number of multiplets in the low-energy theory can be determined from the first cohomology of certain associated bundles which are constructed from the vector bundle, \eqref{Vdef}. This information, together with the various group-theoretical details, is summarised in \tabref{table1}.
\begin{table}
\begin{tabular}{|l|c|c|c|c|c|c|}\hline
\varstr{12.4pt}{7pt} $(SU(2)\times SU(3))_{\bf q}$&$({\bf 1},{\bf 1})_{{\bf e}_a-{\bf e}_b}$&$({\bf 1},{\bf 3})_{-{\bf e}_a-{\bf e}_b}$&
$({\bf 1},\overline{\bf 3})_{{\bf e}_a+{\bf e}_b}$&$({\bf 2},{\bf 3})_{{\bf e}_a}$&$({\bf 2},\overline{\bf 3})_{-{\bf e}_a}$&
$({\bf 2},{\bf 1})_{{\bf e}_a+{\bf e}_b+{\bf e}_c}$\\\hline\hline
\varstr{12.4pt}{7pt} Constraint&&$a\leq b$&$a\leq b$&&&$a\leq b\leq c$\\\hline
\varstr{12.4pt}{7pt} Particle, $F$&$e_{a,b},\;S_{a,b}$&$\tilde{d}_{a,b},\;\tilde{u}_{a,b}$&$d_{a,b},\;u_{a,b}$&$Q_a$&$\widetilde{Q}_a$&$L_{a,b,c},\;H_{a,b,c},\;\bar{H}_{a,b,c}$\\
\hline
\varstr{12.4pt}{7pt} Bundle&$U_a\otimes U_b^*$&$U_a^*\otimes U_b^*$&$U_a\otimes U_b$&$U_a$&$U_a^*$&$U_a\otimes U_b\otimes U_c$\\
\varstr{12.4pt}{7pt} &&$\wedge^2U_a^*$&$\wedge^2U_a$&&&$\wedge^2U_a\otimes U_b$, \\
\varstr{12.4pt}{7pt} &&&&&&$U_a\otimes \wedge^2U_b$, $\wedge^3 U_a$\\\hline
\varstr{12.4pt}{7pt} Contained in&$V\otimes V^*$&$\wedge^2V^*$&$\wedge^2V$&$V$&$V^*$&$\wedge^3V$\\\hline
\varstr{12.4pt}{7pt} $Y(F)$&$y_a-y_b$&$-y_a-y_b$&$y_a+y_b$&$y_a$&$-y_a$&$y_a+y_b+y_c$\\\hline
\varstr{12.4pt}{7pt} $Y_{\rm phys}(F)$&$2,\;0$&$-2/3,\;4/3$&$2/3,\;-4/3$&$1/3$&$-1/3$&$-1,\;-1,\;1$\\\hline
\end{tabular}
\caption{\it Particle content of models with bundle structure group $H$, a unitary split of $SU(6)$. The multiplicity of each type of multiplet is determined by the first cohomology of the associated bundle. The indices $a,b,\dots $ label the summands of the bundle, \eqref{Vdef}, and are in the range $a,b,\ldots =1,\dots ,f$.}\tablabel{table1}
\end{table}
In the last two rows of this table we have also listed the hypercharge of the multiplets in terms of the embedding vector \eqref{ydef} and the physically required hypercharge. Finding hypercharge embeddings ${\bf y}$ and associated patterns of multiplets which do indeed lead to the correct values of hypercharge for all standard model multiplets - and no additional multiplets with exotic charges - is a strong model-building requirement which will be analysed in detail below.

\section{Embedding of hypercharge}\label{yemb}
As explained above, the embedding of hypercharge $U_Y(1)$ into $S(U(1)^f)$ is described by a vector ${\bf y}$ as in \eqref{ydef}. Using the decomposition \eqref{248dec} and \eqref{SU6dec}, the normalization of $U_Y(1)$ can be computed as
\begin{equation}
 \frac{g^2}{g_Y^2}~=~\frac{1}{120}\,{\rm Tr}(Y^2)~=~\frac{1}{2}|{\bf y}^2|~=~\frac{1}{2}\sum_{a=1}^f\,n_a\,y_a^2
\end{equation} 
The standard normalization of hypercharge which is appropriate for gauge unification in its conventional form and which is realized for the usual embedding of hypercharge into $SU(5)$ is given by $g^2/g_Y^2=5/3$. Hence, if we wish to implement the conventional picture of gauge unification, we should require that
\begin{equation}
 \sum_{a=1}^f\,n_a\,y_a^2~\simeq~ \frac{10}{3}\eqlabel{unifcond}
\end{equation} 
is satisfied for our hypercharge embedding, either exactly or to sufficient accuracy.

We can now ask the following, purely group-theoretical question. For which embedding vectors ${\bf y}$ can we assign $S(U(1)^f)$ charges to one  standard model family such that we obtain the correct hypercharge for all multiplets and such that the unification condition \eqref{unifcond} is satisfied? If we require the unification condition~\eqref{unifcond} exactly, it turns out that there is a very limited range of possibilities, which is summarised in \tabref{table2}.
\begin{table}[h]
\begin{tabular}{|c|l|}\hline
\varstr{12.4pt}{7pt} Splitting type ${\bf n}$&Allowed ${\bf y}$ vectors\\\hline\hline
\varstr{12.4pt}{7pt}$(4,1,1)$&\sout{$(1/3,1/3,-5/3)$}\\\hline
\varstr{12.4pt}{7pt}$(3,2,1)$&\sout{$(1/3,1/3,-5/3)$}, $(-2/3,1/3,4/3)$\\\hline
\varstr{12.4pt}{7pt}$(2,2,2)$&no solution\\\hline
\varstr{12.4pt}{7pt}$(3,1,1,1)$&\sout{$(1/3,1/3,1/3,-5/3)$}, $(-2/3,1/3,1/3,4/3)$\\\hline
\varstr{12.4pt}{7pt}$(2,2,1,1)$&\sout{$(1/3,1/3,1/3,-5/3)$}, $(1/3,-2/3,-2/3,4/3)$\\\hline
\varstr{12.4pt}{7pt}$(2,1,1,1,1)$&\sout{$(1/3,1/3,1/3,1/3,-5/3)$}, $(1/3,-2/3,-2/3,-2/3,4/3)$, $(-2/3,-2/3,1/3,1/3,4/3)$\\
\varstr{10pt}{7pt}&$(5/6,-7/6,-2/3,-1/6,1/3)$, $(-5/21,-17/21,-11/21,1/3,31/21)$\\\hline
\varstr{12.4pt}{7pt}$(1,1,1,1,1,1)$&\sout{$(1/3, 1/3, -5/3, 1/3, 1/3, 1/3)$},$(1/3, 4/3, -2/3, -2/3, -2/3, 1/3)$\\
\varstr{10.pt}{7pt}&$(1/3, 5/6, -7/6, -1/6, -2/3, 5/6)$,$(1/3, 7/12, -17/12, 1/12, -5/12, 5/6),\ldots$\\\hline
\end{tabular}
\caption{\it All embedding vectors ${\bf y}$ (modulo re-ordering) for the various splitting patterns \eqref{splits} which can lead to the correct hypercharge for one family and satisfy the unification condition \eqref{unifcond} exactly. The crossed-out vectors correspond to the conventional embedding of hypercharge into $SU(5)$.}\tablabel{table2}
\end{table}

A few remarks about this table are in order. First, note that certain splitting types in \eqref{splits} are excluded right away and, hence, do not appear in the table. Clearly, ${\bf n}=(6)$, which corresponds to a bundle structure group $SU(6)$, is excluded since no $U(1)$ symmetry which could account for hypercharge is left over in this case. Further constraints on the splitting type arise as follows. The first Chern classes of the constituent bundles $U_a$ have to satisfy
\begin{equation}
 \sum_{a=1}^fc_1(U_a)=0\;,\qquad \sum_{a=1}^fy_ac_1(U_a)=0\; , \eqlabel{kconds}
\end{equation}
where the first equation simply states that $c_1(V)=0$ and the second equation is the condition for the hypercharge gauge boson to be massless.\footnote{Note that the second condition in~\eqref{kconds} guarantees that the one-loop contribution to the mass of the hypercharge gauge boson explored in Refs.~\cite{Lukas:1999nh, Blumenhagen:2005ga} vanishes.} Now consider the splitting types ${\bf n}=(n_1,n_2)$ into two summands. In this case, both first Chern classes, $c_1(U_1)$ and $c_1(U_2)$, must  vanish since the conditions \eqref{kconds} are independent. But with $c_1(U_1)=c_1(U_2)=0$ the bundle structure group reduces and is no longer of the type \eqref{Hdef}. Hence, these cases have been discarded in \tabref{table2}. 

\vspace{2pt}
Embedding vectors of the form ${\bf y}=(1/3,\ldots ,1/3,-5/3)$, together with \eqref{kconds}, imply that $c_1(U_f)=0$ and, hence, they also lead to a reduced structure group outside the class specified by \eqref{Hdef}. In fact, these cases correspond to the conventional embedding of hypercharge into an intermediate $SU(5)$ GUT which is then broken by a Wilson line. This is the standard heterotic model-building route which is, of course, perfectly viable. However, in this paper we are focusing on a direct flux breaking to the standard model without any intermediate GUT and, hence, these embedding vectors have been crossed out in \tabref{table2}.

\vspace{2pt}
This leaves us with a fairly limited number of possibilities, one for each of the splitting types ${\bf n}=(3,2,1),\; (3,1,1,1),\;(2,2,1,1)$ and four for the splitting type ${\bf n}=(2,1,1,1,1)$. Only the Abelian case, ${\bf n}=(1,1,1,1,1,1)$, where the vector bundle is a sum of line bundles comes with a large number of possibilities, indicated by the dots in the last row of \tabref{table2}, although with increasingly complicated fractions. In fact, the general solution for the Abelian case can be written in the form
\begin{equation}
{\bf y}=\left(\frac{1}{3},\alpha,\alpha-2,\frac{2}{3} - \alpha,\frac{1}{2} (1-\alpha-s),\frac{1}{2} (1-\alpha+s)\right)\; ,\quad s=\frac{1}{3}\sqrt{-63\alpha^2+114\alpha-31}\eqlabel{allsol}
\end{equation}
where $\alpha$ is a free parameter which should be chosen such that the resulting ${\bf y}$ vector is rational.\\[3mm]
Evidently, the above classification of hypercharge vectors is fairly restrictive. We can slightly relax our requirements by asking the unification condition, \eqref{unifcond}, to be satisfied approximately, within 5\%, rather than exactly and then redo the classification. For simplicity, we will only carry this out for the simplest splitting type, the Abelian splitting into a sum of line bundles with ${\bf n}=(1,1,1,1,1,1)$. To be precise, we ask the following question. For the Abelian splitting type, ${\bf n}=(1,1,1,1,1,1)$, which hypercharge embeddings ${\bf y}$ allow for a pattern of $S(U(1)^6)$ charge assignments such that we obtain the correct hypercharges for all multiplets in one standard model family and the unification condition, \eqref{unifcond}, is satisfied approximately, to within 5\%? The answer to this question is the following four families of embedding vectors, 
\bea\eqlabel{y1234}
\bold y_1 (\alpha, \beta) &= \left(-\frac53,\; \frac13, \;\frac13,\; \frac13 - \alpha, \;\frac13-\beta,\; \frac13+\alpha+\beta\right)\ ,\\
\bold y_2 (\alpha) &=\left(-\frac23,\; \frac13,\; \frac13,\; \frac43,\; -\frac23 - \alpha,\; -\frac23 +\alpha \right) \ , \\
\bold y_3 (\alpha, \beta) &=\left(-\frac53, \;\frac13, \;\frac13 - \alpha, \;\frac13 + \alpha,\; \frac13 - \beta, \;\frac13 + \beta\right) \ , \\
\bold y_4 (\alpha, \beta) &=\left(\frac13, \;-\frac53 - \alpha, \;\frac13 - \alpha, \;\frac13+\alpha,\; \frac13 - \beta + \alpha, \;\frac13 + \beta\right) \ , 
%\bold y_1 (\alpha, \beta) &= \left(-\frac53, \frac13, \frac13, \alpha, 1 -\alpha - \beta, \beta\right)\ , \\
%\bold y_2 (\alpha)&=\left(-\frac23, \frac13, \frac13, \frac43, -\frac43 - \alpha, \alpha\right)\ , \\
%\bold y_3 (\alpha, \beta)&=\left(-\frac53, \frac13, \frac23 - \alpha, \alpha, \frac23 - \beta, \beta\right)\ , \\
%\bold y_4 (\alpha, \beta)&=\left(\frac13, -\frac43 - \alpha, \frac23 - \alpha, \alpha, \frac13 + \alpha - \beta, \beta\right) \ ,  
\eea
which depend on one or two parameters.
Again, this is a fairly restrictive result.\\[3mm]
We end this section with a remark related to gauge unification. Normally, for models with hypercharge flux, one would expect threshold corrections to the gauge kinetic function. These might spoil ``natural" gauge unification just as an incorrect normalization of hypercharge would. In Ref.~\cite{Lukas:1999nh}, Eqs. (5.14) and (5.19), the difference between the $U_Y(1)$ and the $SU_W(2)\times SU_c(3)$ gauge kinetic functions has been calculated as
\begin{equation}
 \delta f_{ab}\sim T^i\,d_{ijk}\,c_1^j(U_a)\,c_1^k(U_b)\; ,\eqlabel{deltaf}
\end{equation}
where $T^i$ are the K\"ahler moduli. In general, this expression is non-vanishing but we have to specialize the gauge field to the hypercharge direction $y^a$. However, from the second condition in \eqref{kconds} which guarantees that hypercharge is massless, we have $y^ac_1^i(U_a)=0$. This implies that the correction~\eqref{deltaf} vanishes if at least one of the gauge fields corresponds to hypercharge. Hence,  there is neither an additional threshold correction for hypercharge nor kinetic mixing of hypercharge with any of the other $U(1)$ symmetries.

To summarise, we conclude that hypercharge can be embedded into $E_8$ in a number of non-standard ways, such that the physically correct hypercharges for the standard model fields can be obtained and ``natural" gauge unification is realised. We emphasise that our viewpoint so far has been purely group-theoretical. In other words, the viable hypercharge embeddings we have found \emph{allow} for patterns of $S(U(1)^f)$ charges such that all values for hypercharge come out correctly. Whether such patterns can actually be realised by an underlying geometry and vector bundle is another question to which we now turn.

\section{A no-go argument}\label{nogo}
Conventionally, heterotic standard models are built based on an ``intermediate" GUT, typically with an $SU(5)$ or $SO(10)$ gauge group, which is subsequently broken by a Wilson line. For such models, hypercharge is of course embedded into the GUT group, in the usual way. This approach requires a Calabi-Yau manifold with a non-trivial first fundamental group or, equivalently, a Calabi-Yau manifold with a freely-acting discrete symmetry, so that a Wilson line can indeed be introduced. Calabi-Yau manifolds with freely-acting discrete symmetries are relatively rare (see, for instance, Refs.~\cite{Braun:2010vc,Batyrev:2005jc}) and, in addition, such discrete symmetries are often not easy to find, so this may be considered a disadvantage of the conventional model-building route. A considerable advantage of this approach is the relative ease with which a physically promising particle spectrum can be obtained. Let us briefly discuss this for the case of an intermediate $SU(5)$ GUT. One standard model family can be grouped into the $SU(5)$ representations ${\bf 10}$ and $\bar{\bf 5}$ and, at the GUT level, the chiral asymmetry of these representations is given by the indices ${\rm ind}(V)$ and ${\rm ind}(\wedge^2 V)$, respectively, where $V$ is the relevant vector bundle with an $SU(5)$ structure group (or a rank-four sub-group thereof). In fact, for $SU(5)$ bundles, these indices are equal, ${\rm ind}(\wedge^2 V)={\rm ind}(V)$. Hence, to obtain a promising model at the GUT level, the only index condition we need to require is ${\rm ind}(V)=-3|\Gamma|$, where $|\Gamma|$ is the order of the freely-acting symmetry group $\Gamma$. Taking the quotient by $\Gamma$ will reduce this to precisely three GUT families and, since the Wilson line does not affect the chiral asymmetry, this will lead to three chiral standard model multiplets of each type. In other words, a promising spectrum with three chiral families is obtained by imposing a single index condition on the vector bundle.

Let us now compare this situation to the one for the models discussed in this paper, where $E_8$ is broken to the standard model directly through flux, without the need for an intermediate GUT and Wilson lines. A clear advantage of pure flux-breaking is that we do not require a non-trivial first fundamental group for the Calabi-Yau manifold - such models can, in principle, be built on any Calabi-Yau manifold. The disadvantage becomes apparent from \tabref{table1}. A physically promising three-family spectrum with the correct values of hypercharge requires satisfying a large number of index conditions on the many bundles in \tabref{table1}, so that all standard model multiplets appear in the right $S(U(1)^f)$ charge sector to produce the correct physical hypercharge for a given embedding ${\bf y}$ and also to avoid the appearance of multiplets with exotic hypercharge. This is to be compared with just one index condition which needs to be imposed on models with underlying GUT symmetry. 

Guided by this observation, we now analyse the index constraints which should be imposed to obtain physically promising spectra in more detail. For now, we do this for the simplest, Abelian split pattern with ${\bf n}=(1,1,1,1,1,1)$ and later extend our results to the general case. 

We begin with a line bundle sum
\begin{equation}
 V=\bigoplus_{a=1}^6L_a \eqlabel{Vlb}
\end{equation}
and also introduce the notation $x_a=c_1(L_a)$ for the first Chern classes of the constituent line bundles. Further, we fix a specific hypercharge embedding ${\bf y}=(y_1,\ldots ,y_6)$, for example, one of the cases found in the previous section. Then, the requirements of $c_1(V)=0$ and vanishing hypercharge mass translate into the conditions
\begin{equation}
 \sum_{a=1}^6x_a=0\;,\qquad \sum_{a=1}^6y_ax_a=0\; . \eqlabel{xlin}
\end{equation}  
A glance at \tabref{table1} indicates the conditions we need to require for a physical spectrum. In order to obtain three chiral families of quarks with the correct hypercharges, we need
\begin{equation}
 \sum_{a:y_a=1/3}{\rm ind}(L_a)=-3\;,\quad \sum_{a<b:y_a+y_b=2/3}{\rm ind}(L_a\otimes L_b)=-3\;,\quad
 \sum_{a<b:y_a+y_b=-4/3}{\rm ind}(L_a\otimes L_b)=-3\; . \eqlabel{gencons}
\end{equation} 
To avoid quarks with exotic hypercharges we also must impose that
\begin{eqnarray}
{\rm ind}(L_a)&=&0\mbox{ if }y_a\neq 1/3\\
{\rm ind}(L_a\otimes L_b)&=&0\mbox{ if }a<b\mbox{ and } y_a+y_b\notin \{2/3,-4/3\}\; .
\end{eqnarray}
Finally, a chiral asymmetry with the wrong sign - which would lead to mirror quarks - should be avoided in each line bundle sector, so we require
\begin{equation}
 -3\leq {\rm ind}(L_a)\leq 0\;,\qquad -3\leq{\rm ind}(L_a\otimes L_b)\leq 0\; ,
\end{equation} 
for all $a,b=1,\ldots, 6$. For the leptons, we cannot impose an overall constraint on the index (the relevant bundles are real and, hence, their index vanishes) but we still need to ensure the absence of leptons with exotic hypercharges. This amounts to the vanishing conditions
\begin{eqnarray}
{\rm ind}(L_a\otimes L_b\otimes L_c)&=&0\mbox{ if }a<b<c\mbox{ and }y_a+y_b+y_c\notin \{-1,1\}\label{v1}\\
{\rm ind}(L_a\otimes L_b^*)&=&0\mbox{ if } y_a-y_b\notin\{-2,0,2\}\; .\eqlabel{v2}
\end{eqnarray} 
How many independent conditions the above equations amount to depends on the structure of the hypercharge embedding ${\bf y}$. As a rule of thumb, the more complicated ${\bf y}$, the more conditions have to be satisfied. 

To see if the above physical conditions can be satisfied, we now express the various indices in terms of the underlying topological data, so that
 \begin{eqnarray}
  {\rm ind}(L_a)&=&\frac{1}{6}x_a^3+\frac{1}{12}x_ac_2(TX)\eqlabel{l1}\\
  {\rm ind}(L_a\otimes L_b)&=&\frac{1}{6}(x_a+x_b)^3+\frac{1}{12}(x_a+x_b)c_2(TX)\eqlabel{l2}\\
  {\rm ind}(L_a\otimes L_b\otimes L_c)&=&\frac{1}{6}(x_a+x_b+x_c)^3+\frac{1}{12}(x_a+x_b+x_c)c_2(TX)\eqlabel{l3}\\
  {\rm ind}(L_a\otimes L_b^*)&=&\frac{1}{6}(x_a-x_b)^3+\frac{1}{12}(x_a-x_b)c_2(TX)\; .\eqlabel{ll}
\end{eqnarray} 
Hence, all relevant line bundle indices depend on the six first Chern classes $x_a$ and the second Chern class of the Calabi-Yau tangent bundle $c_2(TX)$. However, two of the six quantities $x_a$ can be eliminated from these index expressions in favour of the remaining four by using the linear relations, \eqref{xlin}. For definiteness we assume that $x_5$, $x_6$ have been eliminated and that all index expressions are written in terms of $x_\alpha$, where $\alpha=1,\ldots ,4$ and $c_2(TX)$. To further parametrise our ignorance of the underlying geometry we introduce the variables $X_{\alpha\beta\gamma}=x_\alpha x_\beta x_\gamma$ and $Z_\alpha=x_\alpha c_2(TX)$, where $\alpha\leq\beta\leq\gamma$ and $\alpha,\beta,\gamma =1,\ldots ,4$. All index expressions, \eqref{l1}--\eqref{ll}, can then be written as linear functions of the $24$ variables $X_{\alpha\beta\gamma}$ and $Z_\alpha$. Combined with the physical conditions on the indices listed above this leads to a system of linear equations (and inequalities) for $X_{\alpha\beta\gamma}$ and $Z_\alpha$.

We can now ask if this linear system has a solution, for a given hypercharge embedding ${\bf y}$. It turns out that for all possible ${\bf y}$ vectors satisfying the unification condition, \eqref{unifcond}, exactly (as given in the last row of \tabref{table2} or, more generally, by using an arbitrary rational vector of the form \eqref{allsol}),
%with $\alpha=k/l$ and $-100\leq k,l\leq 100$
the answer to this question (by direct computation) is ``no." Hence, remarkably, we have shown that heterotic models with pure flux breaking of a single $E_8$ and exact unification normalization of hypercharge can never lead to a physically acceptable particle spectrum. We emphasise that this conclusion does not rely on any particular Calabi-Yau manifold or class of Calabi-Yau manifolds (since we have absorbed the relevant topological data of the Calabi-Yau manifold into the variables $X_{\alpha\beta\gamma}=x_\alpha x_\beta x_\gamma$ and $Z_\alpha=x_\alpha c_2(TX)$) but is completely general. 

This argument can be repeated for the hypercharge embeddings ${\bf y}$ from \eqref{y1234} which lead to approximate unification. We find that the associated systems of linear equations have no solutions in all cases but one.\footnote{To be precise, for $\bold y$ vectors not in the family,~\eqref{yexcep}, we have confirmed that the Diophantine system for $X_{abc}$ and $Z_a$ has no solutions if the parameters $\alpha$ and $\beta$ in~\eqref{y1234} take the form $k/l$, with $-50\leq k,l\leq 50$, such that the unification condition,~\eqref{unifcond}, is satisfied approximately to within 5\%.}
%To be precise, we confirmed that the Diophantine system for $X_{abc}$ and $Z_a$ has no solutions if $\alpha$ and $\beta$ take the form $k/l$, with $-50\leq k,l\leq 50$, such that the corresponding $\bold y$ in~\eqref{y1234} does not belong to the family,~\eqref{yexcep}, and the unification condition,~\eqref{unifcond}, is satisfied approximately to within 5\%.}
The single remaining case which cannot be excluded in this way is based on the hypercharge embedding
\begin{equation}
 {\bf y}=\left(\frac{1}{3},\;\frac{1}{3},\;\frac{1}{3},\;-\frac{5}{3},\;\frac{1}{3}-\alpha,\;\frac{1}{3}+\alpha\right)\; , \eqlabel{yexcep}
\end{equation}
a common sub-case of the three two-parameter families in \eqref{y1234} with $\alpha \to 0, \beta \to \alpha $ (and suitably re-ordered). Moreover, this case leads to an essentially unique solution for the index conditions which reads
\begin{align}
 {\rm ind}(L_1)={\rm ind}(L_2)={\rm ind}(L_3)=-1\quad&\rightarrow\quad Q_1,\,Q_2,\,Q_3\nn\\
 {\rm ind}(L_1\otimes L_4)={\rm ind}(L_2\otimes L_4)={\rm ind}(L_3\otimes L_4)=-1\quad&\rightarrow\quad u_{1,4},\,u_{2,4},\,u_{3,4}\nn\\
 {\rm ind}(L_5\otimes L_6)=-3 \quad &\rightarrow\quad3\,d_{5,6}\nn\\
 {\rm ind}(L_4\otimes L_5\otimes L_6)=- {\rm ind}(L_1\otimes L_2\otimes L_3)=-3\quad&\rightarrow\quad 3\, L_{4,5,6}\eqlabel{indcons}\\
  {\rm ind}(L_1\otimes L_4^*)={\rm ind}(L_2\otimes L_4^*)={\rm ind}(L_3\otimes L_4^*)=-1\quad&\rightarrow\quad e_{1,4},\,e_{2,4},\,e_{3,4}\nn\\
  {\rm ind}(L_1\otimes L_2^*)=  -{\rm ind}(L_2\otimes L_1^*)=2-X_{112}\quad&\rightarrow\quad S_{1,2}~{\rm or}~S_{2,1} \nn\\
  {\rm ind}(L_1\otimes L_3^*)= -{\rm ind}(L_3\otimes L_1^*)=2+X_{111}+X_{112} \quad&\rightarrow\quad S_{1,3}~{\rm or}~S_{3,1}\nn\\
  {\rm ind}(L_2\otimes L_3^*)=-{\rm ind}(L_3\otimes L_2^*)=6-X_{112}+X_{222}\quad&\rightarrow\quad S_{2,3}~{\rm or}~S_{3,2}\nn
%  {\rm ind}(L_1\otimes L_3^*)=-4-Z_1/2+X_{112}\quad&\rightarrow\quad S_{1,3}\nn\\
%  {\rm ind}(L_2\otimes L_3^*)=-Z_1/2-X_{112}\quad&\rightarrow\quad S_{2,3}\nn
\end{align} 
Finally, we should discuss the other splitting types in \eqref{splits}. Those splitting types require non-Abelian constituent bundles, $U_a$, so that the analogues of the index relations, \eqref{l1}--\eqref{ll}, also depend on higher Chern classes, $c_2(U_a)$, $c_3(U_a)$, in addition to $c_1(U_a)$. However, the splitting principle\footnote{It should be briefly noted here that the splitting principle must be applied with care. It is {\it not} the case that non-Abelian bundles can be split over the Calabi-Yau manifold. Rather, for any bundle $V \to X$ there exists a flag space and a map $s:F(V) \to X$ such that $s^{*}(V)$ decomposes as a direct sum of (complex, not necessarily holomorphic) line bundles and $c_{k}(s^*(V))=s^*(c_k(V))$). We are fortunate that the conditions in \eqref{l1} - \eqref{ll} do not depend explicitly on either holomorphy of the bundle or the Calabi-Yau condition on the base; thus, the no-go results of the Abelian case continue to hold here as well.}\cite{bott_tu} asserts that the total Chern class of $V$ can be expressed as 
%each total Chern class can be written as
\begin{equation}
\prod_{a=1}^{f} c(U_a)=\prod_{a=1}^{f}\prod_{i=1}^{n_a}(1+x_{ai})
\end{equation}
for suitable classes $x_{ai}$ of the second cohomology.  When expressed in terms of $x_{ai}$ the index relations assume precisely the same form as in the Abelian case and the above no-go argument can be applied in the same form, provided the hypercharge embedding ${\bf y}$ in the non-Abelian case is split into an Abelian counterpart accordingly (so that, for example, the embedding ${\bf y}=(-2/3,1/3,4/3)$ for ${\bf n}=(3,2,1)$ becomes ${\bf y}=(-2/3,-2/3,-2/3,1/3,1/3,4/3)$ in the Abelian case). Based on this argument the hypercharge embeddings which satisfy the unification condition exactly, as listed in \tabref{table2}, cannot lead to a standard model spectrum and are, hence, ruled out. Further, all hypercharge  embeddings with approximate unification which split into one of the vectors in~\eqref{y1234} - with the exception of \eqref{yexcep} which we have not excluded - are ruled out on the same grounds. 

We have checked the above no-go argument by constructing explicit models for the Abelian case, based on the splitting type ${\bf n}=(1,1,1,1,1,1)$ and rank-six line bundle sums~\eqref{Vlb}, generalising the model-building approach described in Refs.~\cite{Anderson:2011ns,Anderson:2012yf,Anderson:2013xka}. We have indeed not found a single model, either for hypercharge embeddings with exact or approximate unification, consistent with a standard model spectrum.

In addition, we have carried out a dedicated search for models based on the hypercharge embedding~\eqref{yexcep}, the one case we were not able to exclude from general arguments. The same scanning techniques and Calabi-Yau geometries used in Refs.~\cite{Anderson:2011ns,Anderson:2012yf,Anderson:2013xka} were employed for an extensive search. Unfortunately, no viable models were found for this case either. The problem seems to be one of integrality. It is difficult to satisfy the two conditions~\eqref{xlin} for the ${\bf y}$ vector~\eqref{yexcep}, together with the index conditions~\eqref{indcons} for all $x_a$ being integral -- as required if these quantities are to represent first Chern classes of line bundles. We do not currently know if this problem is general or related to the specific class of models we have studied. 

\section{Conclusion}
In this paper, we have studied $E_8\times E_8$ heterotic Calabi-Yau models based on flux breaking of the visible $E_8$ group down to the standard model group, without an intermediate GUT and Wilson lines. The non-Abelian part of the standard model group has been embedded into $E_8$ via the maximal sub-group $SU_W(2)\times SU_c(3)\times SU(6)\subset E_8$ which leads to the correct $SU_W(2)\times SU_c(3)$ representations required for a standard model spectrum. We have used bundle structure groups $S(U(n_1)\times\cdots \times U(n_f))\subset SU(6)$, where $\sum_{a=1}^fn_a=6$, so that the low-energy gauge group is $SU_W(2)\times SU_c(3)\times S(U(1)^f)$ and hypercharge embedding into $S(U(1)^f)$ is described by a vector ${\bf y}=(y_1,\ldots ,y_f)$. 

We have studied these models in two steps. First, we have considered the purely group-theoretical aspects of model building. In this context, we have classified all hypercharge embeddings ${\bf y}$ which can lead to the correct standard model hypercharges and have the standard normalization required for ``natural" gauge unification. The results are given in \tabref{table2} (for ${\bf y}$ vectors satisfying the unification condition \eqref{unifcond} exactly) and in~\eqref{y1234} (for ${\bf y}$ vectors satisfying the  unification condition approximately).

In a second step, we have then attempted to build explicit models for these hypercharge embeddings. It turns out that obtaining a realistic spectrum in these cases leads to a highly constrained problem, whereby many index conditions have to be imposed on the internal bundle and its various tensor powers. We have shown for all hypercharge embeddings ${\bf y}$ which satisfy the unification condition~\eqref{unifcond} exactly, that these conditions have no solution for any underlying Calabi-Yau manifold and bundle thereon. Further, for the case of approximate unification, we have obtained a similar no-go result for all hypercharge embeddings except for a single case given by \eqref{yexcep}. These results have been checked by explicit model building, based on the approach outlined in Refs.~\cite{Anderson:2011ns,Anderson:2012yf,Anderson:2013xka}.

In summary, we have shown that heterotic $E_8\times E_8$ models with flux breaking of the visible $E_8$ group to the standard model and ``natural" gauge unification can never lead to a realistic particle physics spectrum - barring one marginal case with approximate unification. Intuitively, obtaining the standard model directly requires many topological index conditions, in fact, too many to be satisfied by any underlying geometry and bundle. This result highlights the benefits of heterotic model-building based on an intermediate GUT theory and Wilson line breaking. In this case, a promising spectrum with three chiral families can be obtained by imposing a single index condition at the GUT level while the subsequent Wilson line breaking preserves the chiral asymmetry for each standard model multiplet. 

There are several generalisations and modifications of the models studied in this paper which we have not discussed explicitly. Firstly, it is possible to consider other embeddings of $SU_W(2)\times SU_c(3)$ into $E_8$, although these tend to lead to exotic representations in the branching of the adjoint of $E_8$. Further, it is possible to consider general hypercharge embeddings into $E_8\times E_8$, as has been done in Ref.~\cite{Blumenhagen:2006ux} (see also~\cite{Nibbelink:2009sp, Blaszczyk:2009in, Blaszczyk:2010db}), rather than into a single $E_8$. In this case, the second $E_8$ is not truly hidden and this gives rise to a range of additional model-building problems. For this reason, we have not considered such models but we expect that the approach presented in this paper - and quite possibly some of the no-go results - will extend to these cases.

Finally, we can speculate about the possible relevance of our results for other model-building approaches. Essentially, we have found that the standard model spectrum can be too complicated and fragmented to result from string theory directly. We require the organising principle of an intermediate GUT, broken by a Wilson line in order to preserve chiral asymmetries, for successful model building. Hypercharge flux remains the preferred breaking mechanism in F-theory models, and it is conceivable that similar no-go results (or at least constraints) can be obtained in global F-theory models in this context. We hope to return to this problem in a future publication.

%%%%%%%%%%%%%%%%
%%%%%%%%%%%%%%%%%%%%%%%%%%%%%%%%%%

%%%%%%%%%%%%%%%%%%%%%%%%%%%%%%%%%%%%%%%%%%%%%%%%%%%%%%%%%%%%%%%%%%%%%%%%%%%%%%%%%%%%%%%%%%%%%%%%
\section*{Acknowledgements} 
We would like to thank the organizers of the ``Geometry and Physics of F-Theory'' Workshop at the University of Heidelberg for hospitality during the early stages of this work. We are also grateful to Kang-Sin Choi, James Gray, Xin Gao and Timo Weigand for useful discussions. LA would like to thank the Mainz Institute for Theoretical Physics for hospitality and partial support during the final stages of this work. SJL thanks CERN, and also CERN-Korea Theory Collaboration funded by National Research Foundation (Korea), for hospitality and support.
AC would like to thank the STFC for support during part of the preparation of this paper. AL~is partially supported by the EPSRC network grant EP/l02784X/1 and by the STFC consolidated grant~ST/L000474/1. LA is supported by NSF grant PHY-1417337. The work of SJL is supported in part by NSF grant PHY-1417316.
%%%%%%%%%%%%%%%%%%%%%%%%%%%%%%%%%%%%%%%%%%%%%%%%%%%%%%%%%%%%%%%%%%%%%%%%%%%%%%%%%%%%%%%%%%%%%%%%

%%%%%%%%%%%%%%%%%%%%%%%%%%%%%%%%%%%%%%%%%%%%%%%%%%%%%%%%%%%%%%%%%%%%%%%%%%%%%%%%%%%%%%%%%%%%%%%%

\end{document}